\documentclass[aps,twocolumn]{revtex4-2}
\usepackage[colorlinks]{hyperref}   
\hypersetup{linkcolor=black, anchorcolor=blue, citecolor=blue}  
\usepackage{booktabs}   
\usepackage{amsmath}    
\usepackage{amssymb}    
\usepackage{braket}     
\usepackage{graphicx}   
\usepackage{subcaption}
\usepackage[a4paper, scale=0.8]{geometry}     
\usepackage{titlesec}

\begin{document}

\title{\textbf{Constraining Axion-Like Particles from observations of \\
AGN B2 2234+28A and 3C 454.3}}

\author{Yu-Chong Chen$^{2}$}
\author{Siyu Chen$^{3}$}
\author{Wei-Cong Huang$^{4}$}
\author{Qing Yang$^{1}$}
\email{yangqing@sztu.edu.cn} 

\author{Hong-Hao Zhang$^{3}$}
\email{zhh98@mail.sysu.edu.cn}

\affiliation{$^{1}$\textit{College of Engineering Physics, Shenzhen Technology University, Shenzhen 518118, China}}
\affiliation{$^{2}$\textit{Christ's College, University of Cambridge, Cambridge, United Kingdom}}
\affiliation{$^{3}$\textit{School of Physics, Sun Yat-Sen University, Guangzhou 510275, China}}
\affiliation{$^{4}$\textit{Shenzhen Middle School, Shenzhen, 518001, China}}


\begin{abstract}
Axion-photon oscillation effect provides a possible explanation for the presence of very-high-energy (VHE) $\gamma$-ray signals from distant sources. In this work, we propose a model-dependent method to select possible sources that may give sufficient constraints on the axion parameters. We investigate such effect in the spectra of active galactic nuclei (AGN) B2 2234+28A and 3C 454.3 based on data obtained from Fermi Large Area Telescope (Fermi-LAT) and MAGIC U.L. We utilize the Markov Chain Monte Carlo method to fit the axion parameters, yielding a result of $g_{a\gamma}=3.05^{+0.51}_{-0.31} \times 10^{-11}$ GeV$^{-1}$ for the axion-photon coupling strength and $m_{a}=5.25^{+2.35}_{-2.65} \times 10^{-8} $ eV for the axion mass. We also perform 95\% confidence level (CL) constraints to set an upper limit for $g_{a\gamma}$.
 \\
 \\
\end{abstract}

\maketitle
 
\section{INTRODUCTION}
The hypothetical pseudoscalar boson called axion originate from addressing the strong CP problem in quantum 
chromodynamics (QCD) \cite{peccei1977cp, weinberg1978new, PhysRevLett.40.279, kim1984mu, cheng1988strong, kim2010axions}.  Existence of analogous particles known as axion-like particles (ALPs) are also predicted by theories beyond the Standard Model such as string theory\cite{svrcek2006axions, Jaeckel:2010ni}, and have been considered as potential constituents of dark matter\cite{preskill1983cosmology, dine1983not, abbott1983cosmological, duffy2009axions}.
ALP's coupling with high energy photons leads to ALP-photon oscillation under transverse external magnetic field \cite{raffelt1988mixing,kaplan1985opening,PhysRevLett.51.1415}. Many laboratory experiments are currently seeking ALP's via this effect \cite{graham2015experimental, turner1990windows}, such as CAST \cite{zioutas1999decommissioned}, OSQAR \cite{pugnat2008results, ballou2015new} and PVLAS \cite{PVLAS:2005sku,PVLAS:2008iru}. The conversion effect also suggests the feasibility of probing ALPs via astrophysical methods, i.e., through modifications in $\gamma$-ray spectra of active galactic nuclei (AGN). Specifically, ALP-photon oscillation reduces the photon loss due to annihilation with extragalactic background light (EBL) \cite{finke2010modeling, gilmore2012semi, dwek2013extragalactic, stecker1971cosmic, kneiske2004implications}, leading to high energy photons that have considerably small probabilities of arrival on Earth being observed \cite{de2007evidence, hooper2007detecting, sanchez2009hints, de2008axion, PhysRevD.77.063001, Meyer_2014}. Many prior works have been conducted utilizing such effect to constrain ALP parameters based on spectra of very high energy (VHE) $\gamma$-ray emission of TeV scale from sources such as Gamma Ray Burst 221009A (GRB 221009A) \cite{lhaaso2023very,Gao:2023und}, the Crab Nebula \cite{bi2021axion}, Markarian 421 (Mrk421) \cite{li2021limits, li2022searching, li2024constraints,Li:2024zst}, Mrk 501 and M87 \cite{long2021probing}.  

Despite the robustness of AGN spectrum data, only those spectra that exhibit observable modifications under ALP theory can provide stringent constraints on ALP parameters. It should be noted that while higher energy photons are more likely to undergo ALP oscillation, the detection of such $\gamma$-ray does not necessarily indicate efficient ALP modification. The relatively short distance of low redshift sources may diminish the effect of EBL absorption, allowing the transmission of VHE photons without ALP effects. An example of this is M87, from which photons ranging from GeV to TeV orders are observed \cite{acciari2008observation, abdo2009fermi, neronov2007production}. However, the source cannot provide stringent constraints\cite{long2021probing} because the main mechanism that enables the detection of its VHE  photons is related to its relatively short distance from Earth, rather than any potential ALP effect. Therefore, we develop a source selection method that takes into account both the ALP oscillations at high energy scale and the reduced EBL absorption at low redshifts. This approach allows us to identify potential sources whose photons would indeed experience ALP oscillations during their journey through space under ALP theories and subsequently alter the Spectral energy distribution (SED).

    
Using such method, we target our constraints on $\gamma$-ray observations of Flat Spectrum Radio Quasars (FSRQs) B2 2234+28A (B2A, $z=0.790$) and 3C 454.3 (3C4, $z=0.859$). B2A exhibited significantly increased activity in the GeV energy band in the recent decade \cite{magic2024constraints}, while 3C4 was once one of the brightest $\gamma$-ray sources in the sky \cite{abdo2009early} and photons of TeV scale was recently detected. The amplified $\gamma$-ray signals of both sources were observed by $Fermi$-LAT \cite{ackermann2013first, ajello2016fermi} and MAGIC \cite{sahakyan2021modelling, albert2006observation, aleksic2016major}, covering an energy range of 1 GeV to 10 TeV, sufficient to conduct stringent constraints given the high redshifts of the two sources.
	
The rest of the paper is organized as follows. In Section \ref{PHOTON PROPAGATION}, we briefly introduce the interstellar propagation of photons considering both EBL absorption and ALP-photon oscillation. In Section \ref{Source Selection}, we present our method to select potential sources whose spectrum would exhibit apparent strengthening in the flux of high-energy photons, based on the critical energy for ALP-photon oscillation and its influence on photon propagation probability. In section \ref{Method}, we describe the statistical and programming method used to analyze the observational data. The derived constraints, complemented by a detailed analysis, are presented in Section \ref{Result}. Finally, we summarize our results and conclusions in Section \ref{Conclusion}.
	
\section{PHOTON PROPAGATION}
	\label{PHOTON PROPAGATION}
	In this section we briefly demonstrate the mechanics of ALP-photon oscillation and how it would lead to an increase in the observational flux of high energy photons. 
 
	\subsection{ALP-Photon Oscillation}
	The Lagrangian of ALP-photon coupling is given by:
	\begin{equation}
		\mathcal{L} = -\frac{1}{4}g_{a\gamma}aF_{\mu\upsilon} \tilde{F}^{\mu\upsilon} = g_{a\gamma}a \boldsymbol E \cdot \boldsymbol B
	\end{equation}
	where $F_{\mu\nu}$ and $\tilde{F}^{\mu\nu}$ are the electromagnetic field tensor and its dual, $g_{a\gamma}$ is the coupling constant of the interaction, $a$ is the ALP amplitude, $\boldsymbol E$ and $\boldsymbol B$ are the electric and magnetic field, respectively. We define $\hat{z}$ as the direction of $\gamma$-ray propagation. For highly relativistic axions (i.e., $m_a \ll E$), we apply the short-wavelength approximation to derive the first-order propagation equation \cite{raffelt1988mixing}:
	\begin{equation}
		(E - i\partial_{z} + \mathcal{M})
		\begin{pmatrix}
			A_{x} \\
			A_{y} \\
			a
		\end{pmatrix}
		= 0
	\end{equation}
	where $E$ is the photon energy, $A_{x}$ and $A_{y}$ are the two photon polarization amplitudes. The photon-axion mixing matrix is
	\begin{equation}
		\mathcal{M} = 
		\begin{pmatrix}
			\Delta_{xx} & \Delta{xy} & \frac{g_{a\gamma}}{2}{B_{T}\cos{\theta}} \\
			\Delta_{yx} & \Delta_{yy} & \frac{g_{a\gamma}}{2}B_{T}\sin{\theta} \\
			\frac{g_{a\gamma}}{2}{B_{T}\cos{\theta}} & \frac{g_{a\gamma}}{2}B_{T}\sin{\theta} &
			\Delta_{a}
		\end{pmatrix}
	\end{equation}
	where $B_{T}$ is the transverse magnetic field strength, and $\theta$ is the angle between $\boldsymbol{B}_{T}$ and $\hat x$, which can be set to 0 for simplicity, reducing $\mathcal{M}$ to
	\begin{equation}
		\mathcal{M} = 
		\begin{pmatrix}
			\Delta_{\bot} & \Delta_{R} & 0 \\
			\Delta_{R} & \Delta_{\parallel} & \Delta_{a\gamma} \\
			0 & \Delta_{a\gamma} & \Delta_{a}
		\end{pmatrix}
	\end{equation}
	Following the approximation demonstrated in \cite{mirizzi2008photon}, the Faraday rotation term can be neglected for VHE photons, leading to $\Delta_{R}$ = $0$. Further omitting the Cotton-Mouton effect \cite{rizzo1997cotton, ejlli2019cmb, hochmuth2007effects}, we have $\Delta_{\bot, \parallel}$ = $\Delta_{\rm pl}$ = $-\omega^2_{\rm pl}/2E$, $\omega^2_{\rm pl}$ = $4\pi \alpha n_{e} / m_{e}$, where $\alpha$ is the fine structure constant, and $n_{e}$ is the electron density. Finally, the remaining terms can be numerically calculated as \cite{wang2023axion}
	\begin{equation}
		\label{Deltaagamma}
		\Delta_{a\gamma} \simeq 3.1 \times 10^{-2} \left( \frac{g_{a\gamma}}{2 \times 10^{-11} \rm GeV^{-1}} \right) \left( \frac{B_{T}}{\mu \rm G} \right) \rm kpc^{-1}
	\end{equation}
	\begin{equation}
		\label{Deltaa}
		\Delta_{a} \simeq -7.8 \times 10^{-3} \left( \frac{m_{a}}{10^{-8} \rm eV} \right)^2 \left( \frac{E}{\rm TeV} \right)^{-1} \rm kpc^{-1}
	\end{equation}
	\begin{equation}
		\label{Deltapl}
		\Delta_{\rm pl} \simeq -1.1 \times 10^{-10} \left( \frac{n_{e}}{10^{-3} \rm cm^{3}} \right) \left( \frac{E}{\rm TeV} \right)^{-1} \rm kpc^{-1}
	\end{equation}
	
	\subsection{Magnetic Field Influence}
	To determine the probability of photons converting to axions in the source galaxy and re-converting in the Milky Way, we must determine the magnetic field distribution in the two galaxies. We adopt the Cellular model \cite{grossman2002effects} to approximate the magnetic field structures, assuming that the galaxy is composed of identical small regions of scale $r$, and the magnetic field $\boldsymbol{B}_{T}$ has the same strength throughout the galaxy, while its direction varies randomly across different regions. Consequently, we can simplify the propagation equation to a 2-dimensional form:
	\begin{equation}
		(E - i\partial_{z} + \mathcal{M}_{2})
		\begin{pmatrix}
			A_{2} \\
			a
		\end{pmatrix}
		= 0
	\end{equation}
	where
	\begin{equation}
		\mathcal{M}_{2}=
		\begin{pmatrix}
			\Delta_{\rm pl} & \Delta{a\gamma} \\
			\Delta_{a\gamma} & \Delta_{a}
		\end{pmatrix}
	\end{equation}
	Diagonalize $\mathcal{M}_{2}$ by the rotation angle
	\begin{equation}
		\label{rotation angle}
		\theta = \frac{1}{2}\arctan{\frac{2\Delta_{a\gamma}}{\Delta_{\rm pl}-\Delta_{a}}},
	\end{equation}
	we can get the probability of a photon with an initial state of $A_{2}(0)$ converting to an axion after traveling through one region:
	\begin{equation}
		\label{conversion probability}
		\begin{split}
			P_{0}(\gamma \to a) & = |\braket{A_{2}(0)|a(r)}|^2 \\
			& = \sin^2{(2\theta)}\sin^2{(\Delta_{\rm osc}r/2)} \\
			& = (\Delta_{a\gamma}r)^2\frac{\sin^2{(\Delta_{\rm osc}r/2)}}{(\Delta_{\rm osc}r/2)^2}
		\end{split}
	\end{equation}
	with the oscillation term $\Delta_{\rm osc}^2 = (\Delta_{\rm pl}-\Delta_{a})^2 + 4\Delta_{a\gamma}^2$. Considering the random orientation of magnetic field in each region, the total probability of a photon converting to an axion across a galaxy with scale $L$ much larger than $r$ is \cite{grossman2002effects}
	\begin{equation}
		\label{total conversion probability}
		P_{\gamma \to a} = \frac{1}{3}\left( 1 - e^{-\frac{3P_{0}L}{2r}} \right).
	\end{equation}
	
	\subsection{Spectral Energy Distribution}
	As the $\gamma$-ray enters the intergalactic region, photons that did not convert to axions would be attenuated by annihilation with EBL, while the axions remain unaffected. These axions could convert back to photons in the Milky way, thus offering a way to preserve the VHE photons in the observational flux. The total probability for a photon to be detected is
	\begin{equation}
		\label{transmission probability}
		P_{\gamma \to \gamma} = P_{\gamma \to a}^{\rm S}P_{a \to \gamma}^{\rm G} + (1-P_{\gamma \to \gamma}^{\rm S})e^{-\tau_{\gamma\gamma}},
	\end{equation}
    where the first term accounts for the axion channel, $P_{\gamma \to a}^{\rm S}$ and $P_{a \to \gamma}^{\rm G}$ represent the probability of photon$\to$ALP conversion in the source galaxy and re-conversion to photon in the Milky Way respectively. Note that when the interaction is stable, $P_{a \to \gamma}$ = $2P_{\gamma \to a}$, as can be seen when $P_{\gamma \to a}$ approaches 1/3 when saturated. The second term in Eq.(\ref{transmission probability}) stands for the normal propagation channel for photons, where $e^{-\tau_{\gamma\gamma}}$ is the EBL absorption term, whose value depends on the photon energy $E$ and the source redshift $z$. In this paper, we adopt the model in \cite{finke2022modeling} to accurately account for background light absorption. Qualitatively, $\tau(E, z)$ is positively correlated with $E$ and $z$, revealing that higher energy signals from more distant sources exhibit stronger absorption, thus making the spectrum enhancement of ALP-photon oscillation more apparent for constraint. Here we do not incorporate the term that accounts for the unabsorbed photons converting to ALP before being received as the magnetic field in the reception area is set as $0.5 \rm {\mu G}$, much less significant compared to the souce's field strength of $5 \rm {\mu G}$. 
	
	Finally, to fit the intrinsic spectrum of the sources, we utilize the power-law with exponential cut-off (PLC) \cite{abramowski2013measurement, albert2022long, aharonian2024curvature} model $\phi_{\rm PLC} = \phi_{0}(E/E_{0})^{-\alpha}{\rm exp}(-E/E_{c})$, where $\phi_{0}$ is the normalization factor, $E_{0}$ is the reference energy set at 1 TeV, and $E_{c}$ is the cutoff energy. The spectral energy distribution (SED) received is thus
	\begin{equation}
		\label{SED}
		E^2\frac{{\rm d} N}{{\rm d} E} = P_{\gamma \to \gamma}(E, z)\phi_{\rm PLC}
	\end{equation}
	in unit of TeV ${\rm cm}^{-2}$ ${\rm s}^{-1}$.
	
	\section{SOURCE SELECTION}
	\label{Source Selection}
	
	In order to perform stringent constraint on the ALP parameters, the enhancement on the source's SED due to ALP-photon conversion must be apparent compared to the original flux. In this section, we characterize such modification effect via the theoretical ratio of the transmission probability of VHE photons under the ALP assumption to that without the ALP influence, thereby selecting a preferred redshift and energy band for statistical tests.
	
	\subsection{Critical Energy of Conversion}
	We begin with the critical energy above which the ALP-photon conversions are efficient \cite{mirizzi2009stochastic, long2021probing}: 
	\begin{equation}
		\begin{split}
		\label{critical energy}
		E_{\rm crit} \sim 38 \left( \frac{m_a}{10^{-6} \rm eV} \right)^2 \left( \frac{10^{-5} \rm G}{B} \right) \\
		\times \left( \frac{6.5\times10^{-11} \rm GeV^{-1}}{g_{a\gamma}} \right).
		\end{split}
	\end{equation}
	Once a possible point in the parameter space of $m_a - g_{a\gamma}$ is selected, we can use Eq.(\ref{critical energy}) to determine at which photon energy interval should ALP conversion effect be significant.
	
\begin{figure}
			\centering
			\includegraphics[width=0.45\textwidth]{"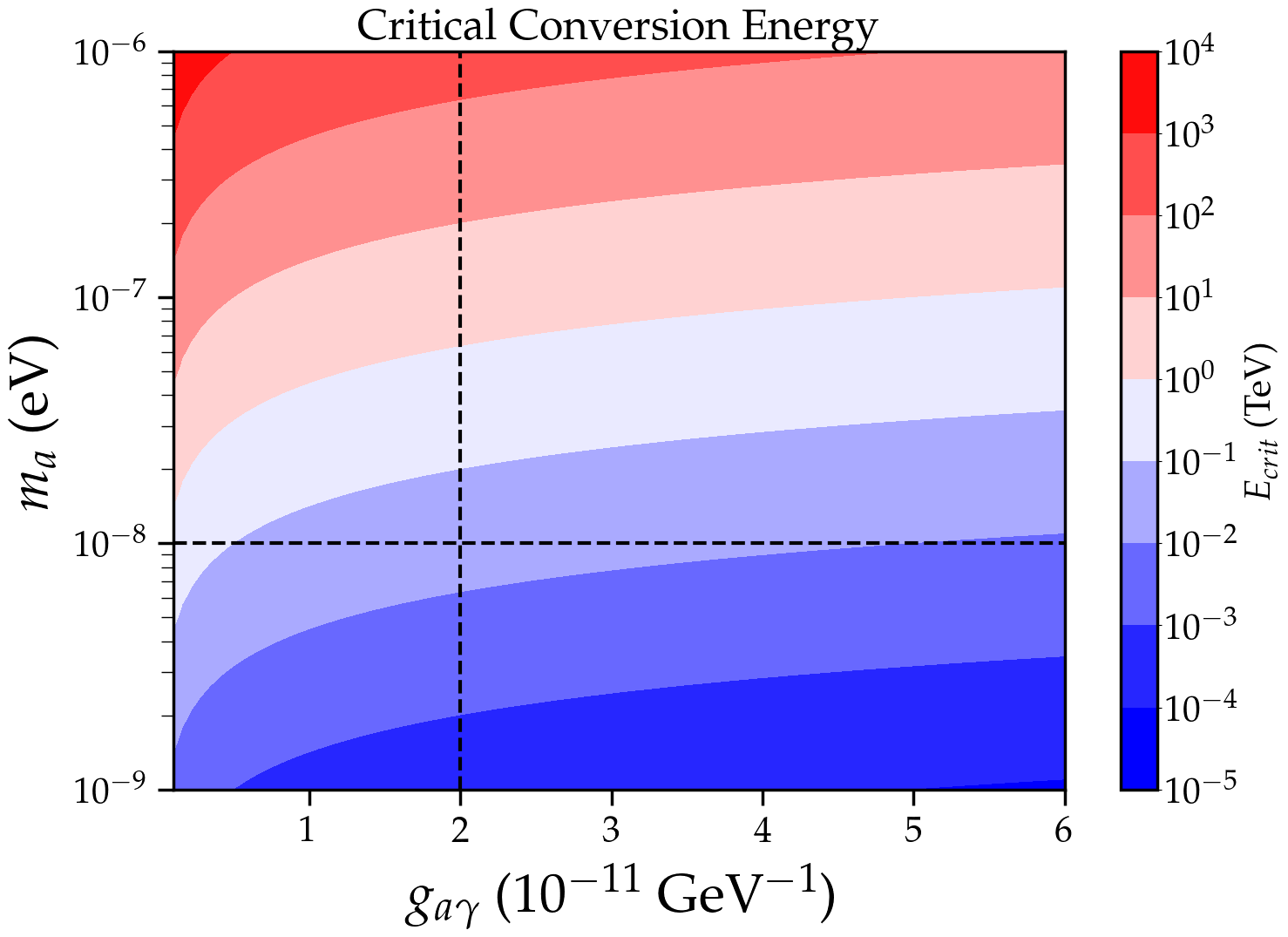"}
			\caption{Contour plot for critical conversion energy against particle mass $m_a$ and coupling strength $g_{a\gamma}$. The two dash lines intersect at ($g_{a\gamma}=2\times10^{-11}\rm GeV^{-1}$, $m_{a}=10^{-8} \rm eV$)}
			\label{fig:critical-energy-graph}
		\end{figure}

	Fig\ref{fig:critical-energy-graph} shows a contour plot for $E_{\rm crit}$ in the parameter space. We choose the point $(g_{a\gamma}=2\times10^{-11}\rm GeV^{-1}$, $m_{a}=10^{-8} \rm eV)$ for reference based on results from \cite{Mastrototaro_2022}. We assume a transverse magnetic field strength of $B_T = 5 \mu \rm G$ and length scales of $L=30 {\rm kpc}$ and $r=3 {\rm kpc}$, which are typical values for AGNs \cite{beck2012magnetic, rodrigues2015galactic}. 
	$E_{\rm crit}$ at this parameter combination falls within the interval of $(0.01, 0.1)$ TeV, indicating that $\gamma$-ray emission below the 0.01 TeV scale cannot yield effective bounds on the parameters if $m_{a} \simeq 10^{-8} \rm eV$. To ensure that efficient photon-ALP oscillation is exhibited, we will require that the $\gamma-$ray signal received from the target source must contain data points with energy higher than 0.1 TeV. Thus, in the remaining of this sectioin, we will focus on the threotical transmission above 0.1 TeV scale.

	\subsection{Optimal Energy and Redshift Interval}
\begin{figure*}
		\subfloat[Transmission Probability Ratio]
		{
			\begin{minipage}{8cm}
				\centering
				\includegraphics[scale=0.4]{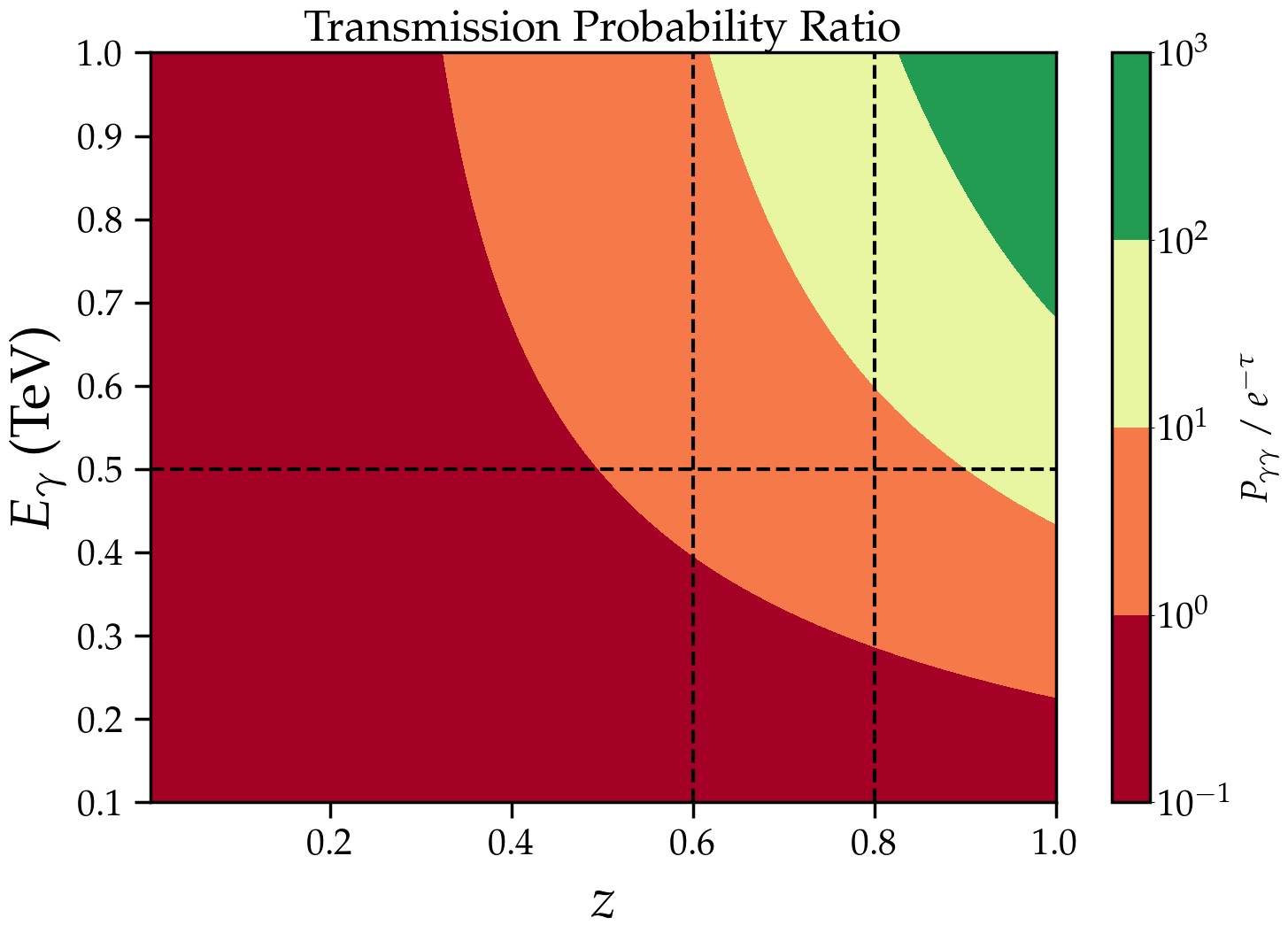}
			\end{minipage}
		}
		\subfloat[Transmission Probability Ratio Extended]
		{
			\begin{minipage}{8cm}
				\centering
				\includegraphics[scale=0.4]{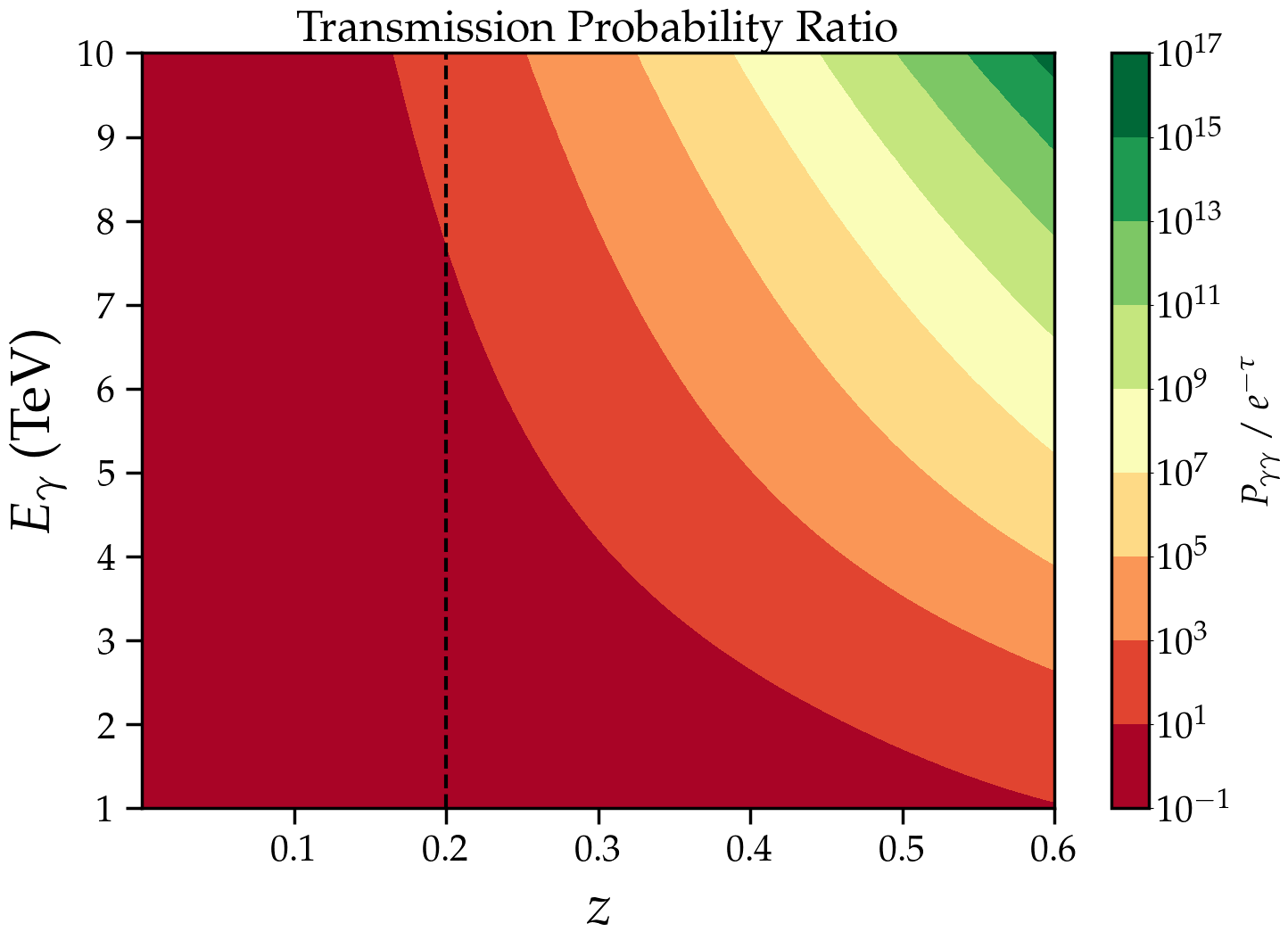}
			\end{minipage}
		}
		\caption{Contour plot for transmission probability ratio $P_{\gamma\gamma}/e^{-\tau}$ against photon energy $E_{\gamma}$ and source redshift $z$. A conservative criteria of 10 is chosen for stringent constraint. A reasonable range is $(0.4, 0.6)$ TeV for $E_{\gamma}$ and $(0.6, 0.8)$ for $z$.}
		\label{fig:transmission probability ratio}
	\end{figure*}

	Without specific information on any particular source, we characterize the influence of ALP oscillation on the SED by comparing VHE photons' transmission probability with and without the ALP effect, based on Eq. (\ref{total conversion probability}) and Eq.(\ref{transmission probability}). Note that in the limit of no conversion, the probability is given by the EBL absorption $e^{-\tau_{\gamma\gamma}}$. Consequently, for a photon with energy $E_{\gamma}$ from a source of redshift $z$, the significance of ALP conversion in its propagation to the Milky Way can be represented in the contour plot in Fig.\ref{fig:transmission probability ratio}. 
	
	For sources with $z < 0.2$ (Fig\ref{fig:transmission probability ratio} (b)), the minimal photon energy for significant ALP enhancement ($P_{\gamma\gamma}/e^{-\tau} \gtrsim 10$ ) in SED would approach $10$ TeV, while for even lower redshift the scale would increase to $\sim 100$ TeV, which is a rather rare case.
	
	Therefore, searching for sources in higher redshift regions in exchange of lowering the energy criteria is an alternative choice to looking for $\sim 10$TeV scale photons from nearby sources. Considering the $0.1$ TeV threshold for conversion, we set $E_{\gamma}=0.5$ TeV as the energy criterion, a conservative choice, to determine the preferred redshift interval at approximately $z \in [0.6, 0.8]$ (Fig\ref{fig:transmission probability ratio} (a)). Note that though sources with redshift higher than this range would theoretically exhibit stronger ALP modification on their SEDs, it is unlikely to observe VHE photons from these regions as the EBL absorption would greatly suppress their spectrum.

\begin{figure}
		\centering
		\includegraphics[width=0.45\textwidth]{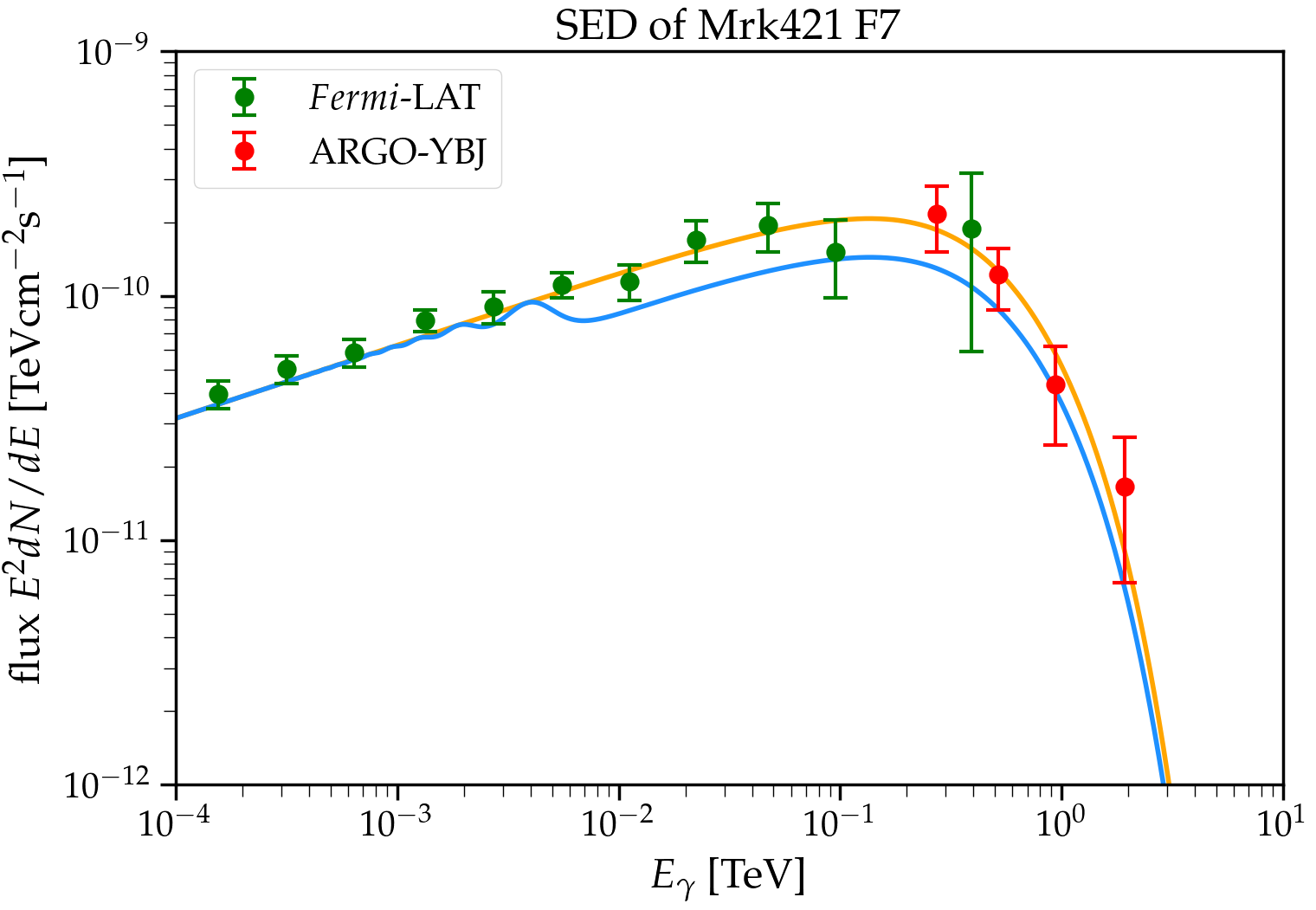}
		\caption{SED of Mrk 421 F7 phase obtained from Astrophysical Radiation with Groundbased Observatory at YangBaJing (ARGO-YBJ) and Fermi-LAT. The blue line represents the scenario without applying ALP theory, while the orange line incorporates the ALP effect with parameters set to $g_{a\gamma}=3\times10^{-11} \rm GeV^{-1}$, $m_{a}=10^{-8} \rm eV$.}
		\label{fig:Mrk421 F7 SED graph}
	\end{figure}

	An instance of failed constraint is provided below with data of Mrk 421 in phase F7 \cite{bartoli20164}. Though photons with energy up to 1 TeV are detected, the low redshift of Mrk 421 ($z=0.031$) means weak EBL absorption during propagation, corresponding to a small ratio in Fig.\ref{fig:transmission probability ratio}, thus mitigating the difference between theories with and without ALP oscillation. As seen in Fig.\ref{fig:Mrk421 F7 SED graph}, there is no significant improvement in the goodness-of-fit by considering ALP theories, and the null theory can explain the flux well without modification. Thus, although the data can still in principle be used to set boundaries for ALP parameter space, the particles' conversion effect is not efficiently at work in this case, and the result is hence not stringent.
	
\begin{figure}
		\centering
		\includegraphics[width=0.45\textwidth]{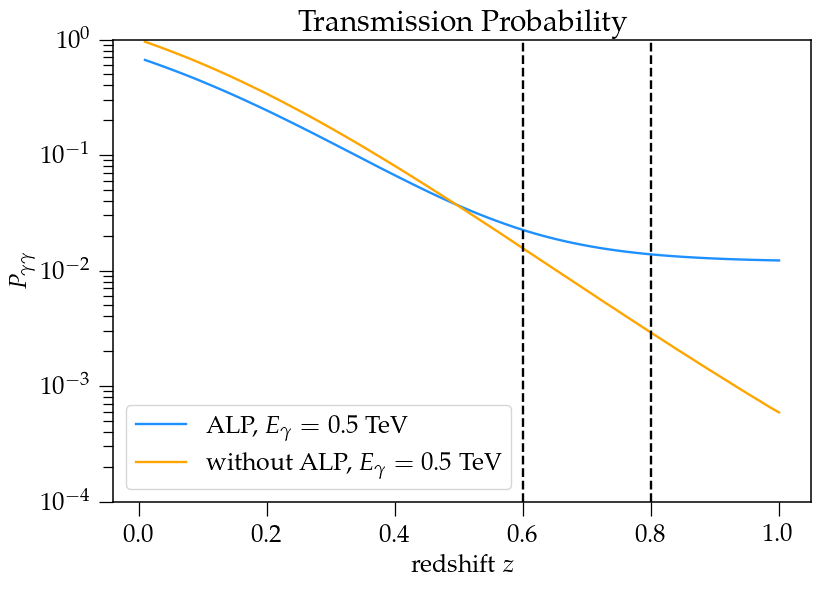}
		\caption{Transmission probability $P_{\gamma\gamma}$ vs. $z$ for photons of $0.5$ TeV. The blue line takes ALP effect into account, while the red line does not. Significant discrepancy begins to occur from $z\sim0.6$.}
		\label{fig:transmission probability line}
	\end{figure}

	Note that $[0.6, 0.8]$ is an approximated redshift region where it is more likely to find sources for effective constraint, but it does not exclude the possibility of successful constraints from other distance ranges. Within this interval, a minimal photon energy of $E_{\rm min} \sim (0.4, 0.6)$ TeV is required for efficient ALP conversion. A more detailed exampled graph is provided in Fig.\ref{fig:transmission probability line}, demonstrating the theoretical transmission probability increase for $0.5$ TeV photons by considering ALP oscillation.
	
	Based on the above threshold, two FSRQs are chosen for our purpose of constraining the ALP parameter space: B2 2234+28A (B2A) and 3C 454.3 (3C4) \cite{rodrigues2024leptohadronic, sikora20083c, li2015multiband}. Due to the lack of data on these two sources' magnetic field distribution, we assume the transverse filed strength to be $B_{T}=5 \mu\rm G$, the region scale to be $r=3 \rm kpc$ \cite{beck2016magnetic, beck2012magnetic, rodrigues2015galactic, wang2023axion}, and divide the total field into 10 such regions ($L=10r$ in Eq.(\ref{total conversion probability})).
	
	\section{METHOD}
	\label{Method}
	\subsection{Markov Chain Monte Carlo}
	\label{MCMC}
	To obtain confidential level distributions for the two parameters, we perform Markov Chain Monte Carlo (MCMC) analysis based on the observed data. For a given source, the $\chi^2$ value is given by
	\begin{equation}
		\label{chi-square}
		\chi^2 = \sum^{N}_{i=1}\frac{(\Phi_i(\phi_{o}, \alpha, E_{c}, g_{a\gamma}, m_a)-\Phi_i^{obs})^2}{\delta_i^2}
	\end{equation}
	where $\Phi_i(\phi_{o}, \alpha, E_{c}, g_{a\gamma}, m_a)$ is the theoretical flux calculated under the ALP effect, $\Phi_i(\rm obs)$ is the observed flux, and $\delta_i$ is the experimental uncertainty. To characterize the goodness-of-fit of a given point in the parameter space, we define the likelihood function
	\begin{equation}
		\label{likelihood function}
		L = {\rm lp} - \frac{1}{2}\chi^2
	\end{equation}
	where the log prior probability $\ln p$ is $-\infty$ for parameter values outside a large given interval and $0$ otherwise. The MCMC simulation would converge to points in the parameter space that generate large $L$. Each step in the simulation is determined by a joint normal distribution centered around each set of data points.
	
	Note that Eq.(\ref{chi-square}) assumes all data points lie within the energy band where ALP oscillation is significant. However, for many sources the majority of data points are within the regime where ALP effect is negligible, Thus using Eq.(\ref{chi-square}) would fail to distinguish the contribution from the intrinsic spectrum parameters and that from ($g_{a\gamma}, m_a$). A  reasonable approach is utilizing low energy data to extrapolate the high energy SED. Hence, it is necessary to first use low energy data points to constrain $(\phi_{0}, \alpha, E_c)$ before determining $(g_{a\gamma}, m_a)$ with the full sample. This provides another reason for not selecting sources with low redshifts: such AGNs are less likely to emit $\gamma$-ray signals with low energy due to insufficient EBL absorption, making the constraint on the intrinsic spectrum parameters ineffective. 
	
	\subsection{CLs Constraint}
	Apart from MCMC simulation, to derive the boundaries of ALP parameter space, we also perform CLs constraints. We aim to find the limit values of $g_{a\gamma}$ and $m_{a}$ above which the predicted spectrum would exceed the observed data. Firstly, we calculate the chi-square value by
	
	\begin{equation}
		\chi^2 = \sum^{N}_{i=1}
	\left\{
		\begin{array}{llr}
			\frac{(\Phi_i^{\rm theo}-\Phi_i^{\rm obs})^2}{\delta_i^2} &, & \Phi_{i}^{\rm theo} \geq \Phi_{i}^{\rm obs} \\
			0 &, &\rm otherwise
		\end{array}
	\right.
	\label{CLs chi-square}
	\end{equation}
	where $\Phi_{i}^{\rm theo}$ is the theoretically calculated flux. For each point in the parameter space $(g_{a\gamma}, m_{a})$, $\Phi^{\rm theo}$ is derived by using the intrinsic spectrum parameter obtained by the MCMC simulations in \ref{MCMC}. 
	Following the method used in \cite{Mastrototaro_2022}, Eq.(\ref{CLs chi-square}) obeys a half-$\chi^2$ distribution. We can then scan the ALP parameter space to exclude regions where $\chi^2/\rm d.o.f > 2.7$, corresponding to a 95\% confidence level.
	
	MCMC simulations are done by the python ensemble toolkit emcee \cite{foreman2013emcee}; calculations on the EBL absorption and PLC model for spectrum are achieved by the Gammapy package \cite{deil2017gammapy, donath2023gammapy} and ebltable package \cite{github.com} respectively. The graph of different CLs constraint results (Fig.\ref{fig:CLcombined}) are generated using codes from \cite{AxionLimits}.
	
	\section{RESULTS}
        \label{Result}
	\subsection{B2A}
\begin{figure}
		\includegraphics[width=0.45\textwidth]{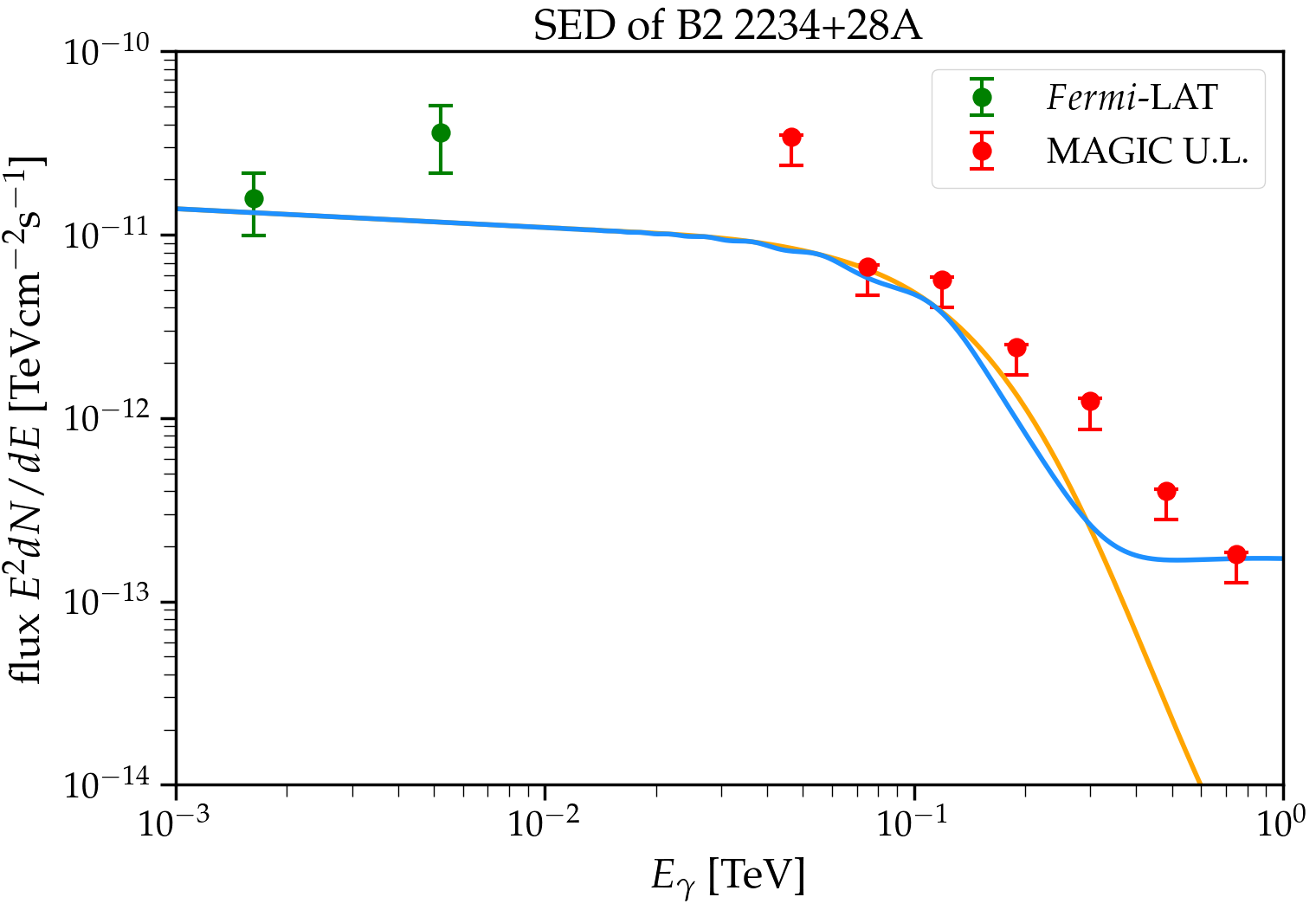}
		\caption{SED of B2 2234+28A. The Green data points are obtained by $Fermi-$LAT, and red data points are from MAGIC. The orange line shows the theoretical spectrum without ALP enhancement, where significant deviation from observational data is observed above $0.1$ TeV. The blue line incorporates the ALP effect with $g_{a\gamma}=3.05\times10^{-11} \rm GeV^{-1}$, $m_{a}=5.25\times10^{-8} \rm eV$, and provides a better fit to the observational data. }
		\label{fig:B2A SED graph}
	\end{figure}

\begin{figure}
		\includegraphics[width=0.45\textwidth]{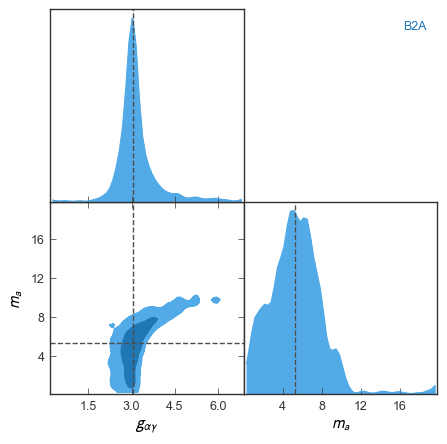}
		\caption{MCMC result of B2 2234+28A. The bottom left panel shows the 68$\%$ and 95$\%$ confidence interval for both parameters. Unit for the scaled coupling strength $g_{a\gamma}$ is $10^{-11} \rm GeV^{-1}$, and that for scaled particle mass $m_{a}$ is $10^{-8} \rm eV$. The narrow range of $g_{a\gamma}$ in this case may serve as a threshold for $g_{a\gamma}$ constraint as the majority of B2A data points are upper limit values.}
		\label{fig:MCMC B2A graph}
	\end{figure}

        For B2A, the first 6 energy bins are not affected by ALP oscillation, so $\phi_{0}, \alpha, E_c$ are first determined with these data points. We can see that while the null hypothesis successfully fits the observational data in the low-energy regime, it predicts a lower flux than what is observed in the high-energy range. This discrepancy can be partially compensated by considering ALP theory, as illustrated in Fig. \ref{fig:B2A SED graph}.
	The MCMC simulation converges at $g_{a\gamma}=3.05^{+0.51}_{-0.31}\times 10^{-11} \rm GeV^{-1}$, $m_a = 5.25^{+2.35}_{-2.65}\times10^{-8}\rm eV$, as is shown in Fig.\ref{fig:MCMC B2A graph}.
	
\begin{figure}
		\centering
		\includegraphics[width=0.45\textwidth]{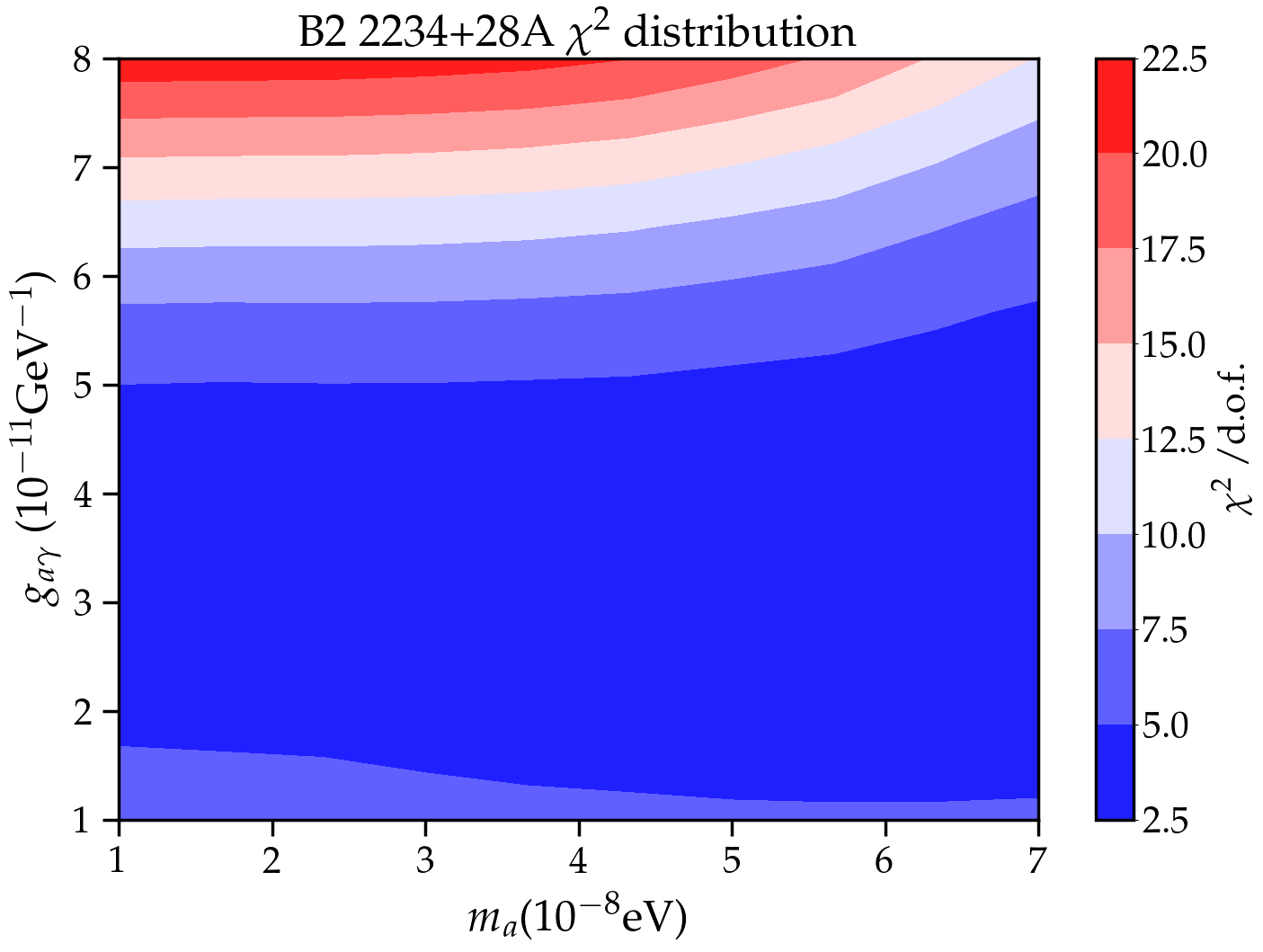}
		\caption{$\chi^2$ contour plot for B2 2234+28A across the $(g_{a\gamma}, m_a)$ parameter plane. A significant region where $g_{a\gamma}$ falls within the range of $[2, 5] \times 10^{-11} \rm{GeV}^{-1}$ can be seen from the plot, corresponding to the peak area in the MCMC simulation in  Fig.\ref{fig:MCMC B2A graph}.}
		\label{B2A chi square graph}
	\end{figure}

	The constraint for $m_{a}$ becomes ineffective at values close to 0. This is because when the ALP mass is small, the photon coupling term $\Delta_{a\gamma} \propto g_{a\gamma}B_T$ (Eq.(\ref{Deltaagamma})) is much larger than $\Delta_{a} \propto m_a^2$ (Eq.(\ref{Deltaa})) , and essentially dominates the influence in the oscillation term $\Delta_{\rm osc}$. It can be shown \cite{gao2024constraints} that the survival probability Eq.(\ref{total conversion probability}) is now $P_{\gamma \to a} \sim g_{a\gamma}^2 B^2 $, indicating that the model is no longer sensitive to ALP mass. This can also be seen via the $\chi^2$ contour plot in Fig.\ref{B2A chi square graph}, where the lines become nearly horizontal as $m_a \to 0$.
	
	It is important to note that the last 7 data points of B2A SED are not actually observed, but rather the upper limit at the given energy. Hence, the result obtained can only serve as an upper threshold for $g_{a\gamma}$. However, B2A remains to be a potential source for effective constraint.
	
	\subsection{3C 454.3}
	\label{3C 454.3}
\begin{figure}
		\includegraphics[width=0.45\textwidth]{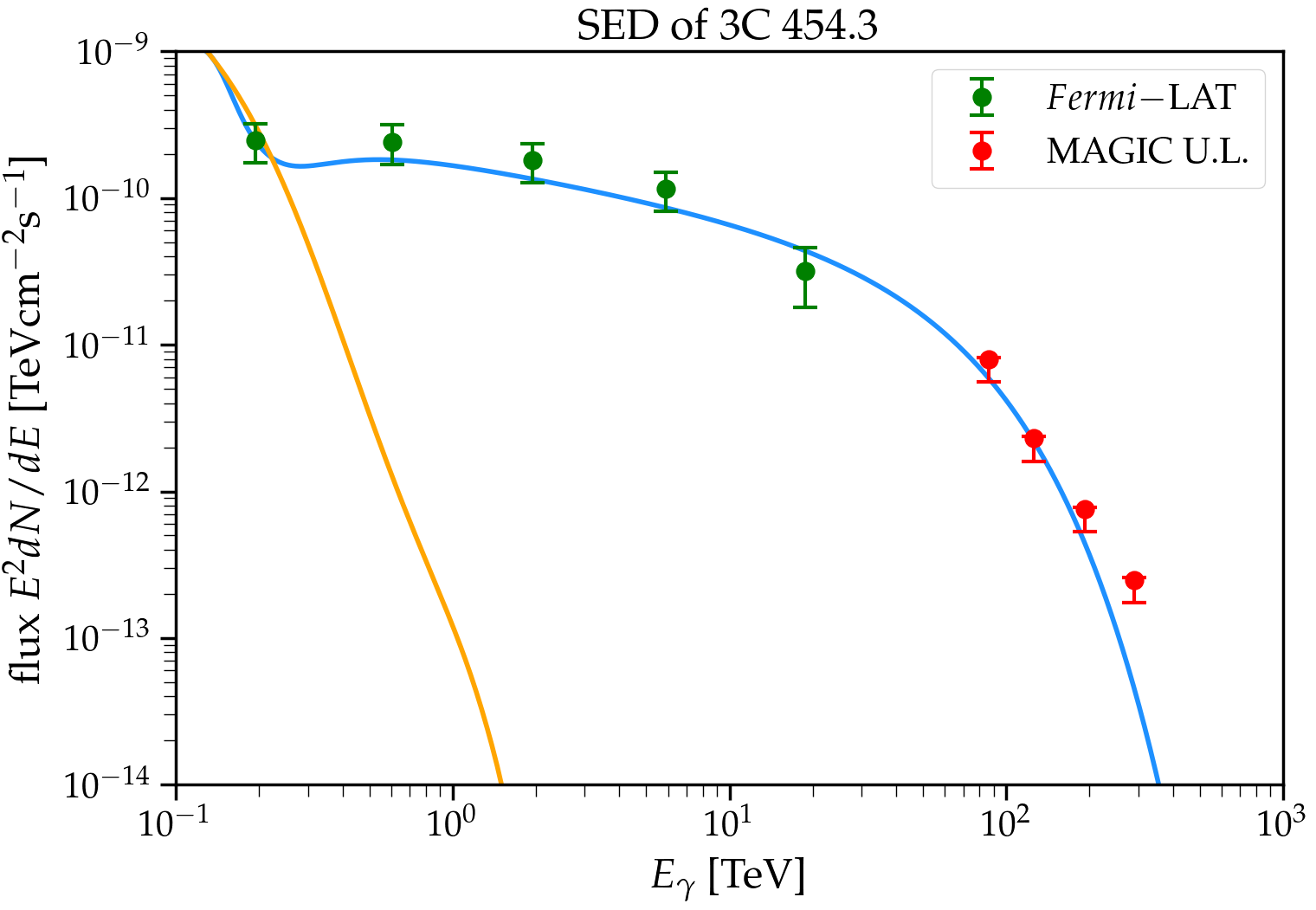}
		\caption{SED of 3C 454.3. Green data points are obtained by $Fermi-$LAT, and red data points are from MAGIC. The orange line shows the theoretical spectrum without ALP enhancement, which falls steeply due to large EBL absorption at VHE regime. The blue line incorporates the ALP effect with $g_{a\gamma}=7.4\times10^{-11} \rm GeV^{-1}$, $m_{a}=5.50\times10^{-8} \rm eV$}
		\label{fig:3C454 SED graph}
	\end{figure}

	The majority of 3C 454.3's data falls in the TeV range, in addition to its high redshift of $z=0.859$, makes it impossible to constrain the parameters purely based on the intrinsic spectrum, as shown in Fig.\ref{fig:3C454 SED graph}.
	The effect of ALP oscillation becomes evident in this scenario, as the high photon energy facilitates efficient conversion, and the strong EBL absorption reduces the VHE photon flux to nearly zero, in stark contrast to the ALP-enhanced line.
	
\begin{figure}

\includegraphics[width=0.45\textwidth]{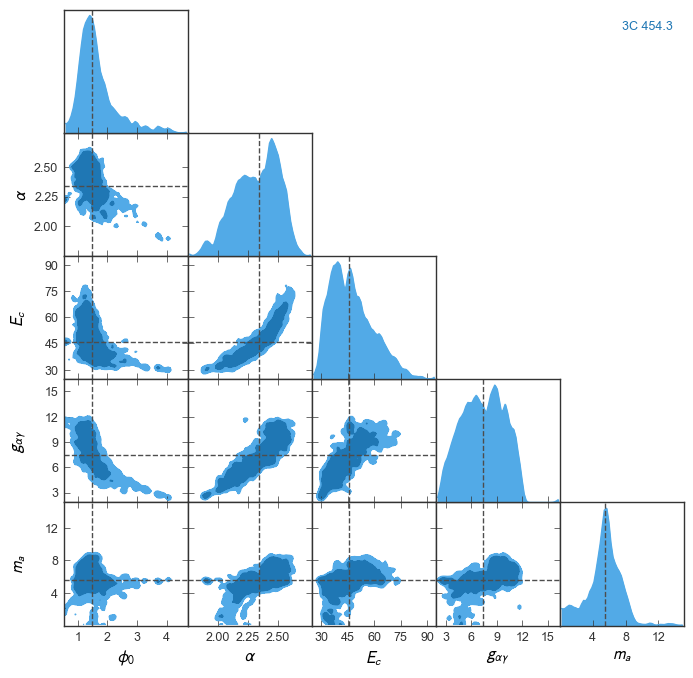}
\caption{MCMC result of 3C 454.3. The first three parameters are from the PLC model, which are the normalization factor $\phi_0$, exponent $\alpha$, and cutoff energy $E_c$, respectively; the last two parameters are the coupling constant and ALP mass as before. Compared to $\phi_0$ and $\alpha$, the constraint on $E_c$ from the simulation is less pronounced because its influence is effectively superseded by the ALP theory. Units of the parameters are identical to those used in table \ref{constraint table}.}
\label{fig:3C454 MCMC}
\end{figure}

	The most stringent constraint is found at $g_{a\gamma} = 7.40^{+2.65}_{-2.74}\times 10^{-11} \rm {GeV^{-1}}$, $m_a=5.50^{+1.69}_{-2.17}\times10^{-8} \rm eV$ (Fig.~\ref{fig:3C454 MCMC}). As shown in Fig.~\ref{fig:3C454 MCMC}, the constraint for cutoff energy $E_c$ is less effective compared to that for the other two parameters in the PLC model. This occurs as both the effects of $E_c$ and ALP theory are to modify the SED curve at energies above a certain threshold. The flux in the high energy regime acquires strong enhancement from $g_{a\gamma}$ (Fig.\ref{fig:3C454 SED graph}), consequently, ALP conversion becomes the dominant factor, while the cutoff influence becomes subtle.

    A summary of the constraint result for both sources can be seen in Table \ref{constraint table}.
	
	\subsection{CLs Exclusion Region}
\begin{figure}

\includegraphics[width=0.45\textwidth]{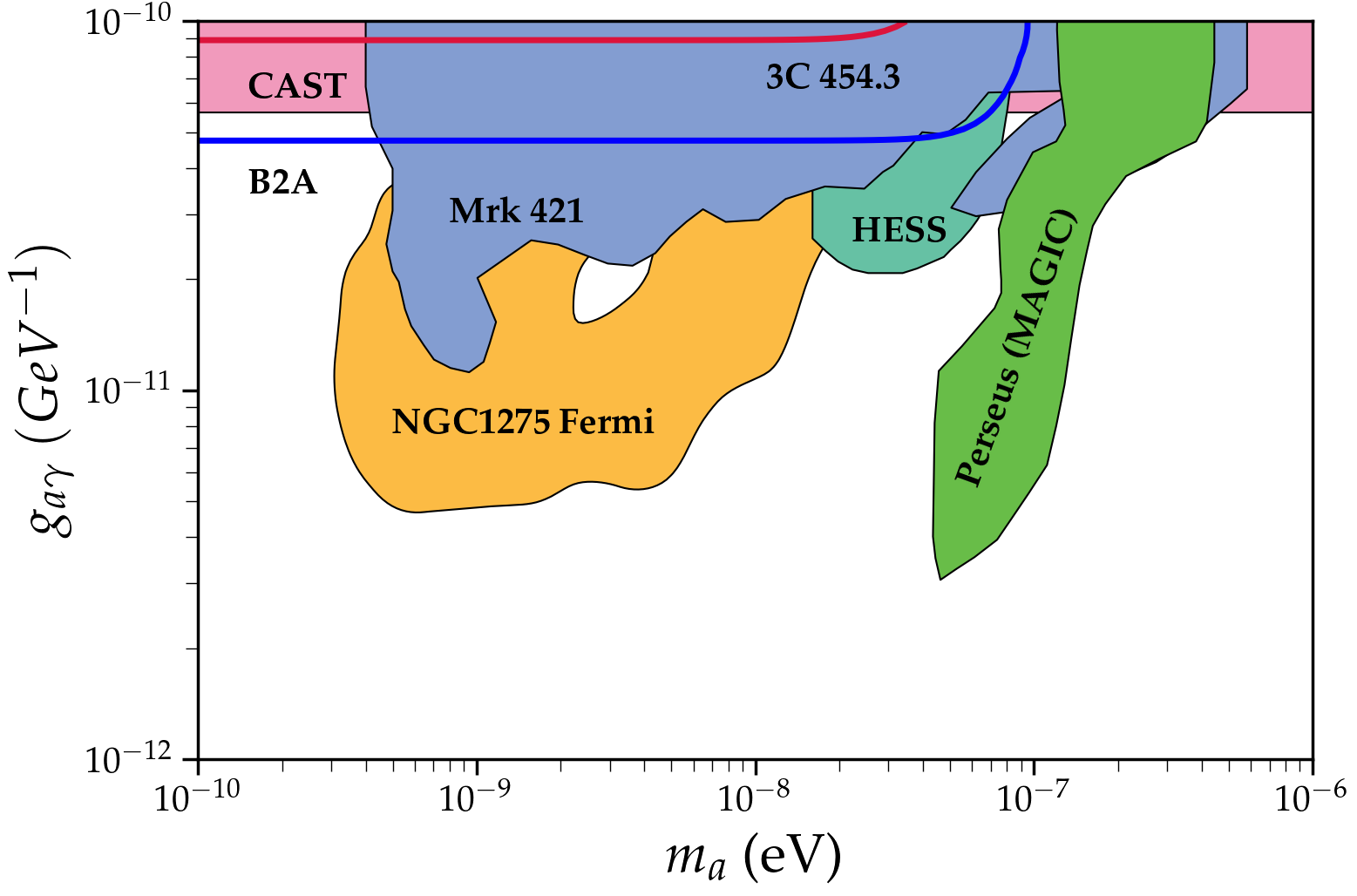}
\caption{Constraints form the CLs method. Space above the blue and red solid lines are the excluded regions derived by this work from 3C 454.3 and B2A respectively. Exclusion region of MAGIC observation on Perseus cluseter of galaxies is also included \cite{aleksic2010magic}}
\label{fig:CLcombined}
\end{figure}

	The result of the CLs constraint is shown in Fig.\ref{fig:CLcombined}, excluding the 95\% confidence level region. We compare our result with the research from CAST \cite{anastassopoulos2017new} ($g_{a\gamma} < 6.6 \times 10^{-11} \rm GeV^{-1}$), HESS \cite{abramowski2013constraints}, Mrk 421 \cite{li2021limits}, and Fermi \cite{ajello2016search}. The region above the blue and red lines are the parameter space excluded by B2A and 3C 454.3, respectively.
	
	Similar to the $\chi^2$ contour plot, the CLs constraints also tend to become independent of the particle mass $m_a$ as it decreases. For both sources, the constraint on the axion mass $m_a$ becomes significant beyond $m_a \approx 10^{-7}, \text{eV}$, as the impact of $m_a$ on the spectral  becomes more pronounced for the CLs method. It is observed that the CLs bound from B2A is, despite the source being less effective in MCMC simulations, more stringent than that from 3C 354.3. This occurs as the photons from B2A lie generally in lower energy regimes, thus requiring less enhancement from ALP oscillation to fit its SED, which in turn leaves less space for $g_{a\gamma}$. Compared to constraints offered by previous works, our result provides additional constraints in the parameter space where $m_{a}$ is small.

\begin{table*}
    \caption{Constraint results from MCMC simulation. The reference energy $E_{0}$ in the PLC model is set at 1 TeV. Significant difference on the coupling strength can be seen, possibly due to B2A lacking actual data in the VHE regime. Constraint on B2A is done first only to the three intrinsic spectrum parameters by low energy data points, before adding in the ALP theory; 3C 454.3 can only yield effective constraint with the ALP parameters to enhance its SED.}
    \begin{ruledtabular}
    \begin{tabular}{ccccccc}
	& \multicolumn{3}{c}{Intrinsic Spectrum} & \multicolumn{2}{c}{ALP Parameters} & \\
	\cmidrule(lr){2-4} \cmidrule(lr){5-6}
	Source & $\phi_{0}$ $({\rm TeV cm^{-2} s^{-1}})$ & $\alpha$ (PLC) & $E_c$ (TeV) & $g_{a\gamma} (10^{-11} \rm GeV^{-1})$ & $m_{a}$ $(10^{-8} \rm eV)$ & $\chi^2 / \rm d.o.f$ \\
	\hline
	B2A & $6.08 \times 10^{-12}$ & 2.13 & 47 & 3.05 & 5.25 & 3.38 \\
	3C 454.3 & $1.49 \times 10^{-9}$ & 2.34 & 46 & 7.40 & 5.50 & 1.82
	\label{constraint table}
    \end{tabular}
    \end{ruledtabular}
\end{table*}

    \section{CONCLUSION}
    \label{Conclusion}
		In this work, we provide a method to identify sources that are capable of imposing stringent constraints on the ALP parameters by calculating the critical energy of ALP-photon conversion and photon propagation probability ratio at different redshifts using the cellular model for galactic magnetic fields. A redshift interval of $z \sim [0.6, 0.8]$ with an energy scale of $0.5 \rm TeV$ are approximated as the appropriate interval. We thus select FSRQs, B2 2234+28A and 3C 454.3, as target sources to perform MCMC simulations and CLs constraint on $g_{a\gamma}$ and $m_{a}$, with data from $Fermi$-LAT and MAGIC. The result from 3C 454.3 yields the best fit value of $g_{a\gamma}= 7.4^{+2.65}_{-2.74}\times 10^{-11} \rm GeV^{-1}$ and $m_a=5.50^{+1.69}_{-2.17}\times10^{-8}$eV, while the result from B2A are $g_{a\gamma}=3.05^{+0.51}_{-0.31} \times 10^{-11}$ GeV$^{-1}$ and $m_{a}=5.25^{+2.35}_{-2.65} \times 10^{-8}$eV. The deviation between the two sets of best-fit values may arise from our assumptions regarding the magnetic field distributions and scales for the two sources. We also use CLs method to establish 95\% CL bounds, revealing improvements in the small $m_{a}$ parameter space.
	
	Further constraints may be achieved by incorporating detailed magnetic field data on the two sources or by accounting for non-uniform field distribution. VHE $\gamma$-ray observations from LHAASO may also provide more feasible and precise constraints on ALPs. We leave these tasks to future works.

\acknowledgments
This work is supported by the National Natural Science Foundation of China (NSFC) under Grant No. 12275367, the Fundamental Research Funds for the Central Universities, and the Sun Yat-sen University Science Foundation.

\bibliographystyle{unsrt}
\bibliography{ConstraintonAxionLikeParticles}

\begin{thebibliography}{10}

\bibitem{peccei1977cp}
Roberto~D Peccei and Helen~R Quinn.
\newblock Cp conservation in the presence of pseudoparticles.
\newblock {\em Physical Review Letters}, 38(25):1440, 1977.

\bibitem{weinberg1978new}
Steven Weinberg.
\newblock A new light boson?
\newblock {\em Physical Review Letters}, 40(4):223, 1978.

\bibitem{PhysRevLett.40.279}
F.~Wilczek.
\newblock Problem of strong $p$ and $t$ invariance in the presence of instantons.
\newblock {\em Phys. Rev. Lett.}, 40:279--282, Jan 1978.

\bibitem{kim1984mu}
Jihn~E Kim and Hans~Peter Nilles.
\newblock The $\mu$-problem and the strong cp-problem.
\newblock {\em Physics Letters B}, 138(1-3):150--154, 1984.

\bibitem{cheng1988strong}
Hai-Yang Cheng.
\newblock The strong cp problem revisited.
\newblock {\em Physics Reports}, 158(1):1--89, 1988.

\bibitem{kim2010axions}
Jihn~E Kim and Gianpaolo Carosi.
\newblock Axions and the strong cp problem.
\newblock {\em Reviews of Modern Physics}, 82(1):557--601, 2010.

\bibitem{svrcek2006axions}
Peter Svrcek and Edward Witten.
\newblock Axions in string theory.
\newblock {\em Journal of High Energy Physics}, 2006(06):051, 2006.

\bibitem{Jaeckel:2010ni}
Joerg Jaeckel and Andreas Ringwald.
\newblock {The Low-Energy Frontier of Particle Physics}.
\newblock {\em Ann. Rev. Nucl. Part. Sci.}, 60:405--437, 2010.

\bibitem{preskill1983cosmology}
John Preskill, Mark~B Wise, and Frank Wilczek.
\newblock Cosmology of the invisible axion.
\newblock {\em Physics Letters B}, 120(1-3):127--132, 1983.

\bibitem{dine1983not}
Michael Dine and Willy Fischler.
\newblock The not-so-harmless axion.
\newblock {\em Physics Letters B}, 120(1-3):137--141, 1983.

\bibitem{abbott1983cosmological}
Laurence~F Abbott and Pierre Sikivie.
\newblock A cosmological bound on the invisible axion.
\newblock {\em Physics Letters B}, 120(1-3):133--136, 1983.

\bibitem{duffy2009axions}
Leanne~D Duffy and Karl Van~Bibber.
\newblock Axions as dark matter particles.
\newblock {\em New Journal of Physics}, 11(10):105008, 2009.

\bibitem{raffelt1988mixing}
Georg Raffelt and Leo Stodolsky.
\newblock Mixing of the photon with low-mass particles.
\newblock {\em Physical Review D}, 37(5):1237, 1988.

\bibitem{kaplan1985opening}
David~B Kaplan.
\newblock Opening the axion window.
\newblock {\em Nuclear Physics B}, 260(1):215--226, 1985.

\bibitem{PhysRevLett.51.1415}
P.~Sikivie.
\newblock Experimental tests of the "invisible" axion.
\newblock {\em Phys. Rev. Lett.}, 51:1415--1417, Oct 1983.

\bibitem{graham2015experimental}
Peter~W Graham, Igor~G Irastorza, Steven~K Lamoreaux, Axel Lindner, and Karl~A van Bibber.
\newblock Experimental searches for the axion and axion-like particles.
\newblock {\em Annual Review of Nuclear and Particle Science}, 65(1):485--514, 2015.

\bibitem{turner1990windows}
Michael~S Turner.
\newblock Windows on the axion.
\newblock {\em Physics Reports}, 197(2):67--97, 1990.

\bibitem{zioutas1999decommissioned}
Konstantinos Zioutas, CE~Aalseth, D~Abriola, FT~Avignone~Iii, RL~Brodzinski, JI~Collar, R~Creswick, DE~Di~Gregorio, H~Farach, AO~Gattone, et~al.
\newblock A decommissioned lhc model magnet as an axion telescope.
\newblock {\em Nuclear Instruments and Methods in Physics Research Section A: Accelerators, Spectrometers, Detectors and Associated Equipment}, 425(3):480--487, 1999.

\bibitem{pugnat2008results}
Pierre Pugnat, Lionel Duvillaret, Remy Jost, Guy Vitrant, Daniele Romanini, Andrzej Siemko, Rafik Ballou, Bernard Barbara, Michael Finger, Miroslav Finger, et~al.
\newblock Results from the osqar photon-regeneration experiment: No light shining through a wall.
\newblock {\em Physical Review D—Particles, Fields, Gravitation, and Cosmology}, 78(9):092003, 2008.

\bibitem{ballou2015new}
R~Ballou, G~Deferne, M~Finger~Jr, M~Finger, L~Flekova, J~Hosek, S~Kunc, K~Macuchova, KA~Meissner, P~Pugnat, et~al.
\newblock New exclusion limits on scalar and pseudoscalar axionlike particles from light shining through a wall.
\newblock {\em Physical Review D}, 92(9):092002, 2015.

\bibitem{PVLAS:2005sku}
E.~Zavattini et~al.
\newblock {Experimental observation of optical rotation generated in vacuum by a magnetic field}.
\newblock {\em Phys. Rev. Lett.}, 96:110406, 2006.
\newblock [Erratum: Phys.Rev.Lett. 99, 129901 (2007)].

\bibitem{PVLAS:2008iru}
M.~Bregant et~al.
\newblock {Limits on Low Energy Photon-Photon Scattering from an Experiment on Magnetic Vacuum Birefringence}.
\newblock {\em Phys. Rev. D}, 78:032006, 2008.

\bibitem{finke2010modeling}
Justin~D Finke, Soebur Razzaque, and Charles~D Dermer.
\newblock Modeling the extragalactic background light from stars and dust.
\newblock {\em The Astrophysical Journal}, 712(1):238, 2010.

\bibitem{gilmore2012semi}
Rudy~C Gilmore, Rachel~S Somerville, Joel~R Primack, and Alberto Dom{\'\i}nguez.
\newblock Semi-analytic modelling of the extragalactic background light and consequences for extragalactic gamma-ray spectra.
\newblock {\em Monthly Notices of the Royal Astronomical Society}, 422(4):3189--3207, 2012.

\bibitem{dwek2013extragalactic}
Eli Dwek and Frank Krennrich.
\newblock The extragalactic background light and the gamma-ray opacity of the universe.
\newblock {\em Astroparticle Physics}, 43:112--133, 2013.

\bibitem{stecker1971cosmic}
Floyd~William Stecker.
\newblock {\em Cosmic gamma rays}, volume 249.
\newblock Scientific and Technical Information Office, National Aeronautics and Space~…, 1971.

\bibitem{kneiske2004implications}
Tanja~M Kneiske, T~Bretz, K~Mannheim, and DH~Hartmann.
\newblock Implications of cosmological gamma-ray absorption-ii. modification of gamma-ray spectra.
\newblock {\em Astronomy \& Astrophysics}, 413(3):807--815, 2004.

\bibitem{de2007evidence}
Alessandro De~Angelis, Marco Roncadelli, and Oriana Mansutti.
\newblock Evidence for a new light spin-zero boson from cosmological gamma-ray propagation?
\newblock {\em Physical Review D—Particles, Fields, Gravitation, and Cosmology}, 76(12):121301, 2007.

\bibitem{hooper2007detecting}
Dan Hooper and Pasquale~D Serpico.
\newblock Detecting axionlike particles with gamma ray telescopes.
\newblock {\em Physical Review Letters}, 99(23):231102, 2007.

\bibitem{sanchez2009hints}
MA~S{\'a}nchez-Conde, D~Paneque, E~Bloom, F~Prada, and A~Dominguez.
\newblock Hints of the existence of axionlike particles from the gamma-ray spectra of cosmological sources.
\newblock {\em Physical Review D—Particles, Fields, Gravitation, and Cosmology}, 79(12):123511, 2009.

\bibitem{de2008axion}
Alessandro De~Angelis, Oriana Mansutti, and Marco Roncadelli.
\newblock Axion-like particles, cosmic magnetic fields and gamma-ray astrophysics.
\newblock {\em Physics Letters B}, 659(5):847--855, 2008.

\bibitem{PhysRevD.77.063001}
Melanie Simet, Dan Hooper, and Pasquale~D. Serpico.
\newblock Milky way as a kiloparsec-scale axionscope.
\newblock {\em Phys. Rev. D}, 77:063001, Mar 2008.

\bibitem{Meyer_2014}
Manuel Meyer, Daniele Montanino, and Jan Conrad.
\newblock On detecting oscillations of gamma rays into axion-like particles in turbulent and coherent magnetic fields.
\newblock {\em Journal of Cosmology and Astroparticle Physics}, 2014(09):003, sep 2014.

\bibitem{lhaaso2023very}
LHAASO collaboration.
\newblock Very high-energy gamma-ray emission beyond 10 tev from grb 221009a.
\newblock {\em Science Advances}, 9(46):eadj2778, 2023.

\bibitem{Gao:2023und}
Lin-Qing Gao, Xiao-Jun Bi, Jun Li, Run-Min Yao, and Peng-Fei Yin.
\newblock {Constraints on axion-like particles from the observation of GRB 221009A by LHAASO}.
\newblock {\em JCAP}, 01:026, 2024.

\bibitem{bi2021axion}
Xiaojun Bi, Yu~Gao, Junguang Guo, Nick Houston, Tianjun Li, Fangzhou Xu, and Xin Zhang.
\newblock Axion and dark photon limits from crab nebula high-energy gamma rays.
\newblock {\em Physical Review D}, 103(4):043018, 2021.

\bibitem{li2021limits}
Hai-Jun Li, Jun-Guang Guo, Xiao-Jun Bi, Su-Jie Lin, and Peng-Fei Yin.
\newblock Limits on axionlike particles from mrk 421 with 4.5-year period observations by argo-ybj and fermi-lat.
\newblock {\em Physical Review D}, 103(8):083003, 2021.

\bibitem{li2022searching}
Hai-Jun Li, Xiao-Jun Bi, and Peng-Fei Yin.
\newblock Searching for axion-like particles with the blazar observations of magic and fermi-lat.
\newblock {\em Chinese Physics C}, 46(8):085105, 2022.

\bibitem{li2024constraints}
Jun Li, Xiao-Jun Bi, Lin-Qing Gao, Xiaoyuan Huang, Run-Min Yao, and Peng-Fei Yin.
\newblock Constraints on axion-like particles from the observation of galactic sources by the lhaaso.
\newblock {\em Chinese Physics C}, 48(6):065107, 2024.

\bibitem{Li:2024zst}
Hai-Jun Li, Wei Chao, and Yu-Feng Zhou.
\newblock {Upper limit on the axion-photon coupling from Markarian 421}.
\newblock 6 2024.

\bibitem{long2021probing}
Guangbo Long, Siyu Chen, Shuo Xu, and Hong-Hao Zhang.
\newblock Probing $\mu$ ev alps with future lhaaso observations of agn $\gamma$-ray spectra.
\newblock {\em Physical Review D}, 104(8):083014, 2021.

\bibitem{acciari2008observation}
VA~Acciari, M~Beilicke, G~Blaylock, SM~Bradbury, JH~Buckley, V~Bugaev, Y~Butt, O~Celik, A~Cesarini, L~Ciupik, et~al.
\newblock Observation of gamma-ray emission from the galaxy m87 above 250 gev with veritas.
\newblock {\em The Astrophysical Journal}, 679(1):397, 2008.

\bibitem{abdo2009fermi}
A.~A. Abdo, M.~Ackermann, M.~Ajello, W.~B. Atwood, M.~Axelsson, L.~Baldini, J.~Ballet, G.~Barbiellini, D.~Bastieri, K.~Bechtol, R.~Bellazzini, B.~Berenji, R.~D. Blandford, E.~D. Bloom, Bonamente, and Others.
\newblock Fermi large area telescope gamma-ray detection of the radio galaxy m87.
\newblock {\em The Astrophysical Journal}, 707(1):55–60, November 2009.

\bibitem{neronov2007production}
Andrii Neronov and Felix~A Aharonian.
\newblock Production of tev gamma radiation in the vicinity of the supermassive black hole in the giant radio galaxy m87.
\newblock {\em The Astrophysical Journal}, 671(1):85, 2007.

\bibitem{magic2024constraints}
MAGIC Collaboration et~al.
\newblock Constraints on vhe gamma-ray emission of flat spectrum radio quasars with the magic telescopes.
\newblock {\em arXiv preprint arXiv:2403.13713}, 2024.

\bibitem{abdo2009early}
A.~A. Abdo, Markus Ackermann, Marco Ajello, WB~Atwood, Magnus Axelsson, Luca Baldini, Jean Ballet, Guido Barbiellini, Denis Bastieri, M~Battelino, et~al.
\newblock Early fermi gamma-ray space telescope observations of the quasar 3c 454.3.
\newblock {\em The Astrophysical Journal}, 699(1):817, 2009.

\bibitem{ackermann2013first}
Markus Ackermann, Marco Ajello, A~Allafort, WB~Atwood, Luca Baldini, Jean Ballet, Guido Barbiellini, Denis Bastieri, K~Bechtol, Aea Belfiore, et~al.
\newblock The first fermi-lat catalog of sources above 10 gev.
\newblock {\em The Astrophysical Journal Supplement Series}, 209(2):34, 2013.

\bibitem{ajello2016fermi}
M~Ajello, Andrea Albert, WB~Atwood, G~Barbiellini, Denis Bastieri, K~Bechtol, Ronaldo Bellazzini, E~Bissaldi, RD~Blandford, ED~Bloom, et~al.
\newblock Fermi-lat observations of high-energy $\gamma$-ray emission toward the galactic center.
\newblock {\em The Astrophysical Journal}, 819(1):44, 2016.

\bibitem{sahakyan2021modelling}
N~Sahakyan.
\newblock Modelling the broad-band emission of 3c 454.3.
\newblock {\em Monthly Notices of the Royal Astronomical Society}, 504(4):5074--5086, 2021.

\bibitem{albert2006observation}
J~Albert, E~Aliu, H~Anderhub, P~Antoranz, A~Armada, M~Asensio, C~Baixeras, JA~Barrio, M~Bartelt, H~Bartko, et~al.
\newblock Observation of gamma rays from the galactic center with the magic telescope.
\newblock {\em The Astrophysical Journal}, 638(2):L101, 2006.

\bibitem{aleksic2016major}
Jelena Aleksi{\'c}, Stefano Ansoldi, Lucio~Angelo Antonelli, P~Antoranz, A~Babic, P~Bangale, M~Barcel{\'o}, JA~Barrio, J~Becerra Gonz{\'a}lez, W~Bednarek, et~al.
\newblock The major upgrade of the magic telescopes, part ii: A performance study using observations of the crab nebula.
\newblock {\em Astroparticle Physics}, 72:76--94, 2016.

\bibitem{mirizzi2008photon}
Alessandro Mirizzi, Georg~G Raffelt1, and Pasquale~D Serpico.
\newblock Photon-axion conversion in intergalactic magnetic fields and cosmological consequences.
\newblock In {\em Axions: Theory, Cosmology, and Experimental Searches}, pages 115--134. Springer, 2008.

\bibitem{rizzo1997cotton}
Carlo Rizzo, Antonio Rizzo, and David~M Bishop.
\newblock The cotton-mouton effect in gases: experiment and theory.
\newblock {\em International Reviews in Physical Chemistry}, 16(1):81--111, 1997.

\bibitem{ejlli2019cmb}
Damian Ejlli.
\newblock On the cmb circular polarization: I. the cotton--mouton effect.
\newblock {\em The European Physical Journal C}, 79(3):1--27, 2019.

\bibitem{hochmuth2007effects}
Kathrin~A Hochmuth and G{\"u}nter Sigl.
\newblock Effects of axion-photon mixing on gamma-ray spectra from magnetized astrophysical sources.
\newblock {\em Physical Review D—Particles, Fields, Gravitation, and Cosmology}, 76(12):123011, 2007.

\bibitem{wang2023axion}
Luohan Wang and Bo-Qiang Ma.
\newblock Axion-photon conversion of grb221009a.
\newblock {\em Physical Review D}, 108(2):023002, 2023.

\bibitem{grossman2002effects}
Yuval Grossman, Sourov Roy, and Jure Zupan.
\newblock Effects of initial axion production and photon--axion oscillation on type ia supernova dimming.
\newblock {\em Physics Letters B}, 543(1-2):23--28, 2002.

\bibitem{finke2022modeling}
Justin~D Finke, Marco Ajello, Alberto Dom{\'\i}nguez, Abhishek Desai, Dieter~H Hartmann, Vaidehi~S Paliya, and Alberto Saldana-Lopez.
\newblock Modeling the extragalactic background light and the cosmic star formation history.
\newblock {\em The Astrophysical Journal}, 941(1):33, 2022.

\bibitem{abramowski2013measurement}
A~Abramowski, Fabio Acero, F~Aharonian, AG~Akhperjanian, G~Anton, S~Balenderan, Agn{\`e}s Balzer, Anna Barnacka, Yvonne Becherini, J~Becker Tjus, et~al.
\newblock Measurement of the extragalactic background light imprint on the spectra of the brightest blazars observed with hess.
\newblock {\em Astronomy \& Astrophysics}, 550:A4, 2013.

\bibitem{albert2022long}
A~Albert, R~Alfaro, C~Alvarez, JR~Angeles Camacho, JC~Arteaga-Vel{\'a}zquez, KP~Arunbabu, D~Avila Rojas, HA~Ayala Solares, V~Baghmanyan, E~Belmont-Moreno, et~al.
\newblock Long-term spectra of the blazars mrk 421 and mrk 501 at tev energies seen by hawc.
\newblock {\em The Astrophysical Journal}, 929(2):125, 2022.

\bibitem{aharonian2024curvature}
F~Aharonian, F~Ait Benkhali, J~Aschersleben, H~Ashkar, M~Backes, V~Barbosa Martins, R~Batzofin, Y~Becherini, D~Berge, K~Bernl{\"o}hr, et~al.
\newblock Curvature in the very-high energy gamma-ray spectrum of m 87.
\newblock {\em Astronomy \& Astrophysics}, 685:A96, 2024.

\bibitem{mirizzi2009stochastic}
Alessandro Mirizzi and Daniele Montanino.
\newblock Stochastic conversions of tev photons into axion-like particles in extragalactic magnetic fields.
\newblock {\em Journal of Cosmology and Astroparticle Physics}, 2009(12):004, 2009.

\bibitem{Mastrototaro_2022}
Leonardo Mastrototaro, Pierluca Carenza, Marco Chianese, Damiano F.~G. Fiorillo, Gennaro Miele, Alessandro Mirizzi, and Daniele Montanino.
\newblock Constraining axion-like particles with the diffuse gamma-ray flux measured by the large high altitude air shower observatory.
\newblock {\em The European Physical Journal C}, 82(11), November 2022.

\bibitem{beck2012magnetic}
Rainer Beck.
\newblock Magnetic fields in galaxies.
\newblock {\em Space science reviews}, 166(1):215--230, 2012.

\bibitem{rodrigues2015galactic}
Luiz Felippe~S Rodrigues, Anvar Shukurov, Andrew Fletcher, and CM~Baugh.
\newblock Galactic magnetic fields and hierarchical galaxy formation.
\newblock {\em Monthly Notices of the Royal Astronomical Society}, 450(4):3472--3489, 2015.

\bibitem{bartoli20164}
Bruno Bartoli, Paolo Bernardini, XJ~Bi, Z~Cao, Sergio Catalanotti, SZ~Chen, TL~Chen, SW~Cui, BZ~Dai, Antonio D’Amone, et~al.
\newblock 4.5 years of multi-wavelength observations of mrk 421 during the argo-ybj and fermi common operation time.
\newblock {\em The Astrophysical Journal Supplement Series}, 222(1):6, 2016.

\bibitem{rodrigues2024leptohadronic}
Xavier Rodrigues, Vaidehi~S Paliya, Simone Garrappa, Anastasiia Omeliukh, Anna Franckowiak, and Walter Winter.
\newblock Leptohadronic multi-messenger modeling of 324 gamma-ray blazars.
\newblock {\em Astronomy \& Astrophysics}, 681:A119, 2024.

\bibitem{sikora20083c}
Marek Sikora, Rafa Moderski, and Greg~M Madejski.
\newblock 3c 454.3 reveals the structure and physics of its “blazar zone”.
\newblock {\em The Astrophysical Journal}, 675(1):71, 2008.

\bibitem{li2015multiband}
HZ~Li, LE~Chen, TF~Yi, YG~Jiang, X~Chen, LZ~L{\"u}, and KY~Li.
\newblock Multiband variability analysis of 3c 454.3 and implications for the center structure.
\newblock {\em Publications of the Astronomical Society of the Pacific}, 127(947):1, 2015.

\bibitem{beck2016magnetic}
Rainer Beck.
\newblock Magnetic fields in spiral galaxies.
\newblock {\em The Astronomy and Astrophysics Review}, 24(1):4, 2016.

\bibitem{foreman2013emcee}
Daniel Foreman-Mackey, David~W Hogg, Dustin Lang, and Jonathan Goodman.
\newblock emcee: the mcmc hammer.
\newblock {\em Publications of the Astronomical Society of the Pacific}, 125(925):306, 2013.

\bibitem{deil2017gammapy}
Christoph Deil, Roberta Zanin, Julien Lefaucheur, Catherine Boisson, Bruno Kh{\'e}lifi, R{\'e}gis Terrier, Matthew Wood, Lars Mohrmann, Nachiketa Chakraborty, Jason Watson, et~al.
\newblock Gammapy-a prototype for the cta science tools.
\newblock {\em arXiv preprint arXiv:1709.01751}, 2017.

\bibitem{donath2023gammapy}
Axel Donath, R{\'e}gis Terrier, Quentin Remy, Atreyee Sinha, Cosimo Nigro, Fabio Pintore, Bruno Kh{\'e}lifi, Laura Olivera-Nieto, Jose~Enrique Ruiz, Kai Br{\"u}gge, et~al.
\newblock Gammapy: A python package for gamma-ray astronomy.
\newblock {\em Astronomy \& Astrophysics}, 678:A157, 2023.

\bibitem{github.com}
Manuel Meyer.
\newblock me-manu/ebltable: ebltable: Python packages to read in and interpolate tables of the extragalactic background light.
\newblock \url{https://github.com/me-manu/ebltable}.
\newblock Accessed November 10, 2022.

\bibitem{AxionLimits}
Ciaran O'Hare.
\newblock cajohare/axionlimits: Axionlimits.
\newblock \url{https://cajohare.github.io/AxionLimits/}, July 2020.

\bibitem{gao2024constraints}
Lin-Qing Gao, Xiao-Jun Bi, Jun Li, Run-Min Yao, and Peng-Fei Yin.
\newblock Constraints on axion-like particles from the observation of grb 221009a by lhaaso.
\newblock {\em Journal of Cosmology and Astroparticle Physics}, 2024(01):026, 2024.

\bibitem{anastassopoulos2017new}
V~Anastassopoulos, S~Aune, K~Barth, A~Belov, G~Cantatore, JM~Carmona, JF~Castel, SA~Cetin, F~Christensen, JI~Collar, et~al.
\newblock New cast limit on the axion-photon interaction.
\newblock {\em arXiv preprint arXiv:1705.02290}, 2017.

\bibitem{abramowski2013constraints}
Attila Abramowski, Fabio Acero, F~Aharonian, Faical Ait~Benkhali, AG~Akhperjanian, E~Ang{\"u}ner, Gisela Anton, Shangkari Balenderan, Agn{\`e}s Balzer, Anna Barnacka, et~al.
\newblock Constraints on axionlike particles with hess from the irregularity<? format?> of the pks 2155-304 energy spectrum.
\newblock {\em Physical Review D—Particles, Fields, Gravitation, and Cosmology}, 88(10):102003, 2013.

\bibitem{ajello2016search}
M~Ajello, A~Albert, Brandon Anderson, Luca Baldini, G~Barbiellini, Denis Bastieri, R~Bellazzini, E~Bissaldi, RD~Blandford, ED~Bloom, et~al.
\newblock Search for spectral irregularities due to photon--axionlike-particle oscillations with the fermi large area telescope.
\newblock {\em Physical Review Letters}, 116(16):161101, 2016.

\bibitem{aleksic2010magic}
Jelena Aleksi{\'c}, LA~Antonelli, P~Antoranz, M~Backes, C~Baixeras, S~Balestra, JA~Barrio, Denis Bastieri, J~Becerra Gonz{\'a}lez, W~Bednarek, et~al.
\newblock Magic gamma-ray telescope observation of the perseus cluster of galaxies: Implications for cosmic rays, dark matter, and ngc 1275.
\newblock {\em The Astrophysical Journal}, 710(1):634, 2010.

\end{thebibliography}
\end{document}